\documentclass[aps,pre,10pt, amsmath]{revtex4-1}

\newcommand{\beq}{\begin{equation}}
\newcommand{\eeq}{\end{equation}}
\newcommand{\half}{\frac{1}{2}}

\newcommand{\mf}{mean-field}
\newcommand{\mft}{mean-field theory}
\newcommand{\fc}{fully connected}
\newcommand{\nn}{nearest-neighbor}
\newcommand{\tcn}{T_{c,N}}
\newcommand{\tci}{T_{c,\infty}}
\newcommand{\e}{\epsilon}
\newcommand{\en}{\epsilon_N} 
\newcommand{\ei}{\epsilon_{\rm eff}}
\newcommand{\im}{Ising model}
\newcommand{\fcim}{fully connected Ising model}
\newcommand{\mt}{\widetilde m}
\newcommand{\mbar}{\overline{m}}
\newcommand{\mc}{Monte Carlo}

\newcommand{\ms}{m_s}
\renewcommand{\dh}{\Delta h}

\newcommand{\psit}{\widetilde \psi}
\newcommand{\psibar}{\overline{\psi}}
\newcommand{\hs}{h_s}
\newcommand{\f}{z}
\newcommand{\fss}{finite size scaling}

\usepackage{graphicx}
\usepackage{subfigure}
\usepackage{verbatim}
\usepackage{color}
\begin{document}

\title{Anomalous mean-field behavior of the \fcim}

\author{Louis Colonna-Romano}
\email{lcolonnaromano@clarku.edu}
\affiliation{Department of Physics, Clark University, Worcester, Massachusetts 01610}

\author{Harvey Gould}
\email{hgould@clarku.edu}
\affiliation{Department of Physics, Clark University, Worcester, Massachusetts 01610}
\affiliation{Department of Physics, Boston University, Boston, Massachusetts 02215}
\author{W. Klein}
\email{klein@bu.edu}
\affiliation{Department of Physics, Boston University, Boston, Massachusetts 02215}
\affiliation{Center for Computational Science, Boston University, Boston, Massachusetts 02215}

\begin{abstract}
Although the \fcim\ does not have a length scale, we show that the critical exponents for thermodynamic quantities such as the mean magnetization and the susceptibility can be obtained using \fss\ with the scaling variable equal to $N$, the number of spins. Surprisingly, the mean value and the most probable value of the magnetization are found to scale differently with $N$ at the critical temperature of the infinite system, and the magnetization probability distribution is not a Gaussian, even for large $N$. Similar results inconsistent with the usual understanding of \mft\ are found at the spinodal. We relate these results to the breakdown of hyperscaling and show that hyperscaling can be restored by increasing $N$ while holding the Ginzburg parameter rather than the temperature fixed, or by doing \fss\ at the pseudocritical temperature where the susceptibility is a maximum for a given value of $N$. We conclude that \fss\ for the \fcim\ yields different results depending on how the \mf\ limit is approached.
\end{abstract}

\maketitle

\section{Introduction}

Mean-field approaches to phase transitions are useful for several reasons. Two of the most important are that they provide a simple way of understanding the nature of critical phenomena~\cite{stanley}, and they are good approximations for systems with long-range interactions and for systems with large molecules~\cite{bigKlein, binder}. Despite the work of Kac and collaborators~\cite{kac}, who defined the applicability of \mf\ theories in a mathematically precise manner, \mf\ approximations are still approached in different ways. These different approaches can be confusing because they can produce different results for the same system. A common approach is to assume that the probability distribution of the order parameter is a Gaussian. Another common approach is to consider a system at its upper critical dimension.

In this paper we investigate another often used approach of understanding \mf\ systems and compare this approach to other ways of doing \mf\ theory. We consider the \fcim\ for which every spin interacts with every other spin. The Hamiltonian of the \fcim\ is given by~\cite{amit, fcim0, bertin, fcim1}
\beq
H = -J_N \!\sum_{i > j,\, j=1}^N \sigma_i \sigma_j - h \sum_{i=1}^N\sigma_i,
\eeq
where $\sigma_i = \pm 1$ and $h$ represents the external magnetic field. The interaction strength $J_N$ is rescaled so that the total interaction energy of a given spin remains the same as $N$ is changed. We take
\beq
J_N = \frac{qJ}{N-1}, \label{JN}
\eeq
with $q=4$. This choice of $q$ yields the \mf\ critical temperature $\tci = 4$, the value of the critical temperature for a square lattice in the limit $N \to \infty$. We have chosen units such that $J/k = 1$, with $k$ equal to the Boltzmann constant. The \fcim\ is sometimes referred to as the ``\mf''~\cite{griffiths}, ``infinitely coordinated''~\cite{botet, botet2}, or ``infinite range''~\cite{zeros} \im.

The standard approach to finite size scaling yields numerical values of the critical exponents by determining how various quantities change with the linear dimension $L$ at the critical temperature of the infinite system~\cite{privman, cardy, barber}. The finite size scaling relations for the \im\ with finite-range interactions include
\begin{align}
\mbar & \sim L^{-\beta/\nu} \\
\chi & \sim L^{\gamma/\nu},
\end{align}
where $\mbar=\overline{|M|}/N$, $|M|$ is the absolute value of the magnetization of the system, the overbar denotes the ensemble average, $\chi$ is the susceptibility per spin, $N$ is the number of spins, and $\beta$, $\gamma$, and $\nu$ are the usual critical exponents~\cite{stanley}. The exponents at the \mf\ critical point are given by
\beq
\gamma = 1, \mbox{ } \beta = 1/2, \mbox{ and } \nu=1/2, \label{mfcpexp}
\eeq
which yields $\mbar \sim L^{-1}$ and $\chi \sim L^2$ if we assume the system can be described by \mft\ at or above the upper critical dimension.

Because the \fcim\ has no length scale, the linear dimension $L$ is not defined. 
One simple way to determine how $\mbar$ and $\chi$ change with $N$ at the critical temperature is to assume that its critical exponents are the same as the \nn\ \im\ in four dimensions, the upper critical dimension~\cite{ellis}. Given this assumption we can write $N \sim L^4$, and hence~\cite{botetnote}
\begin{align}
\mbar & \sim N^{-1/4}. \label{m} \\
\chi & \sim N^{1/2}. \label{chi}
\end{align}
We stress that we will not assume that $N \sim L^4$ to obtain any of our results in the following, and we make this assumption here only to motivate our investigation and simply note that this assumption is only one way of doing \fss\ for \mf\ systems.

As pointed out in Refs.~\cite{defects} and \cite{4d}, the properties of the \im\ in four dimensions and the predictions of other approaches are not always the same. Hence, it is desirable to determine the \fss\ behavior of various properties of the \fcim\ directly. We will find that \fss\ at the critical temperature of the infinite \fcim\ yields results that are inconsistent with both the assumption of a Gaussian probability distribution and several results at the upper critical dimension.

\section{\label{anal}Numerical results for the mean magnetization and the susceptibility}

\begin{figure}[t!]
\centering
\includegraphics[scale=0.6]{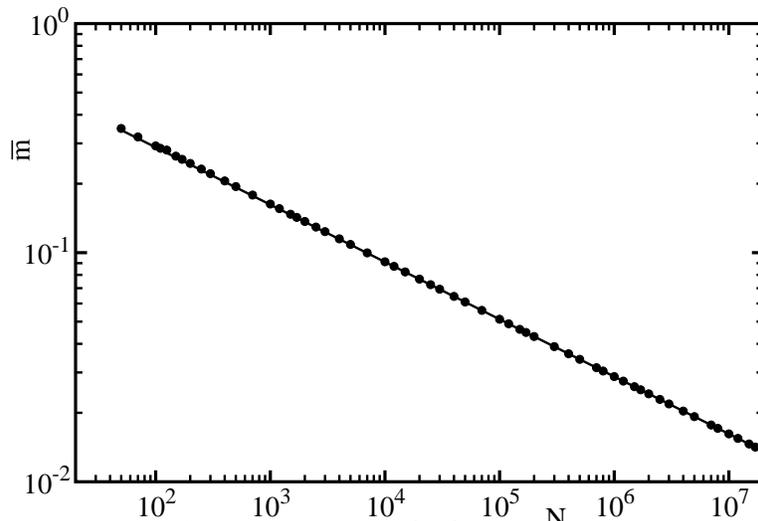}
\vspace{-0.6cm}
\caption{\label{fig:m}Log-log plot of $\mbar$, the mean value of the absolute magnetization per spin, versus $N$, the number of spins, at the critical temperature of the infinite \fcim, $\tci=4$ computed using the exact density of states in Eq.~\eqref{g(M)}. The slope from a least squares fit to $\mbar$ for $10^5\leq N \leq 2\times10^7$ is $-0.2502$, which is consistent with Eq.~\eqref{m}.}
\end{figure}

The exact density of states $g(M)$ of the \fcim\ is given by
\beq
g(M) = \frac{N!}{n!(N-n)!}, \label{g(M)}
\eeq
where $n=(N+M)/2$ is the number of up spins. The probability that the system has magnetization $M$ is proportional to
\beq
P(M) = g(M) e^{-E/T}, \label{P(M)}
\eeq
with the energy $E$ given by
\beq
E = \frac{J_N}{2}(N-M^2) - h M. \label{E}
\eeq
Note that the density of states depends only on $M$. We will refer to $P(M)$ in Eq.~\eqref{P(M)} as a probability, although $P(M)$ is not normalized.

We can evaluate $\chi$ and $m$ numerically as a function of $N$ using the exact density of states in Eq.~\eqref{g(M)}. The only numerical limitation is associated with the rapid increase of $g(M)$ with increasing $N$. Our calculations for $N \leq 2 \times 10^6$ use infinite precision integer arithmetic. Five thousand digits were retained for $2 \times 10^6 < N \leq 2 \times 10^7$. The two numerical approaches give consistent results for $N=2 \times 10^6$.

\begin{figure}[t]
\centering
\includegraphics[scale=0.6]{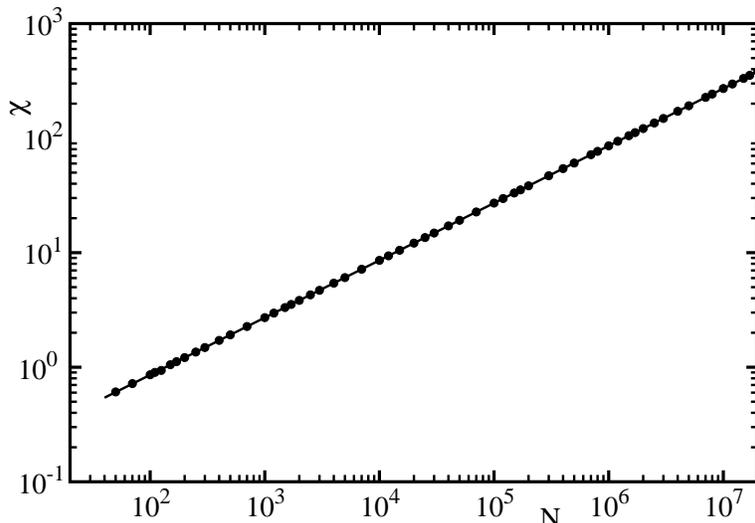}
\vspace{-0.5cm}
\caption{\label{fig:chi}Log-log plot of $\chi$, the susceptibility per spin, versus $N$ at $T=\tci$ for $N \leq 2 \times 10^7$ computed using the exact density of states in Eq.~\eqref{g(M)}. The slope from a least squares fit to $\chi$ for $10^5\leq N \leq 2\times10^7$ is 0.5000, consistent with Eq.~\eqref{chi}.}
\end{figure}

Our numerical results for the $N$ dependence of $\mbar$ and $\chi$ are shown in Figs.~\ref{fig:m} and \ref{fig:chi}, respectively, and are consistent with Eqs.~\eqref{m} and \eqref{chi}.

\section{\label{sec:probable}Most probable value of the magnetization}

We can derive analytical expressions for the $N$-dependence of various quantities using the exact density of states. The usual treatment of the \fcim\ is based on determining the value of $M$ that maximizes $P(M)$. If we use Stirling's formula, $\ln x! \approx x \ln x - x$, we find for large $N$ that
\beq
\frac{d\ln P(M)}{dM} \approx \half \ln\!\frac{(N-n)}{n} + \beta (qJ M + h) = 0. \label{p=0}
\eeq
Equation~\eqref{p=0} yields the usual \mf\ result $m = \tanh \beta(q J m + h)$.

To find the $N$-dependence of $M$ at $T=\tci$ we keep the next term in Stirling's formula, $\ln x! \approx x \ln x - x + \ln \sqrt{2 \pi x}$,
so that $d\ln x!/dx \approx \ln x + 1/2x$.
In this approximation we obtain
\beq
\frac{d\ln P(M)}{dM} \approx \half \ln \frac{1-m}{1+m} + \frac{m}{N}\frac{1}{1-m^2} + \frac{\beta q J m }{1 -1/N} + \beta h= 0. \label{first}
\eeq
We let $h=0$ and keep terms to first-order in $1/N$ and third-order in $m$. The result is
\beq
-m - \frac{m^3}{3} + \frac{m}{N}(1 + m^2) + \beta q J m \Big(1 + \frac{1}{N}\Big)=0. \label{eq:approx}
\eeq
For $ \beta q J = 1$ ($T=\tci$), several terms cancel, and we obtain~\cite{shore}
\beq
m^2 \sim \frac{6}{N} \qquad (N \gg 1). \label{scaling of m2}\\
\eeq

We see from Eq.~\eqref{scaling of m2} that $m \sim N^{-1/2}$, in apparent contradiction with Eq.~\eqref{m}. However, the variable $m$ in Eq.~\eqref{scaling of m2} is the most probable value of the magnetization rather than its mean value. Hence, the mean value and the most probable value of the magnetization scale differently with $N$ at $T=\tci$, behavior that is inconsistent with our usual understanding of \mf.

\begin{figure}[htb]
\centering
\includegraphics[scale=0.6]{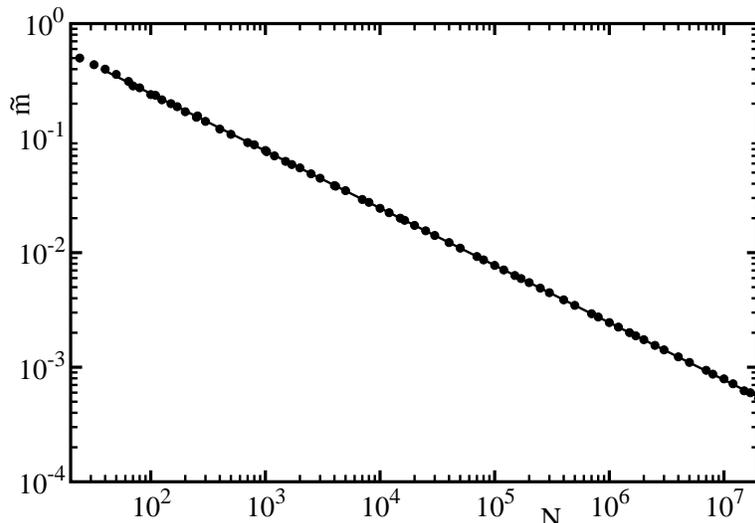}
\vspace{-0.5cm}
\caption{\label{probable}The $N$-dependence of $\mt$, the most probable value of $m$, at $T=\tci$ as determined from Eqs.~\eqref{g(M)}--\eqref{E}. The slope from a least squares fit to $\mt$ for $10^5\leq N \leq 2\times10^7$ is $-0.4997$, consistent with the $N$-dependence of the analytical result in Eq.~\eqref{scaling of m2}.}
\end{figure}

In Fig.~\ref{probable} we plot the $N$-dependence of $\mt$, the most probable (positive) value of $m$, as determined numerically from Eqs.~\eqref{g(M)}--\eqref{E}. We see that the $N$-dependence of $\mt$ is consistent with
\beq
\mt \sim N^{-1/2} \qquad \mbox{(most probable value at $T=\tci$)}.
\eeq

\section{The probability distribution}

\begin{figure}[t]
\centering
\subfigure[]{\includegraphics[scale=0.6]{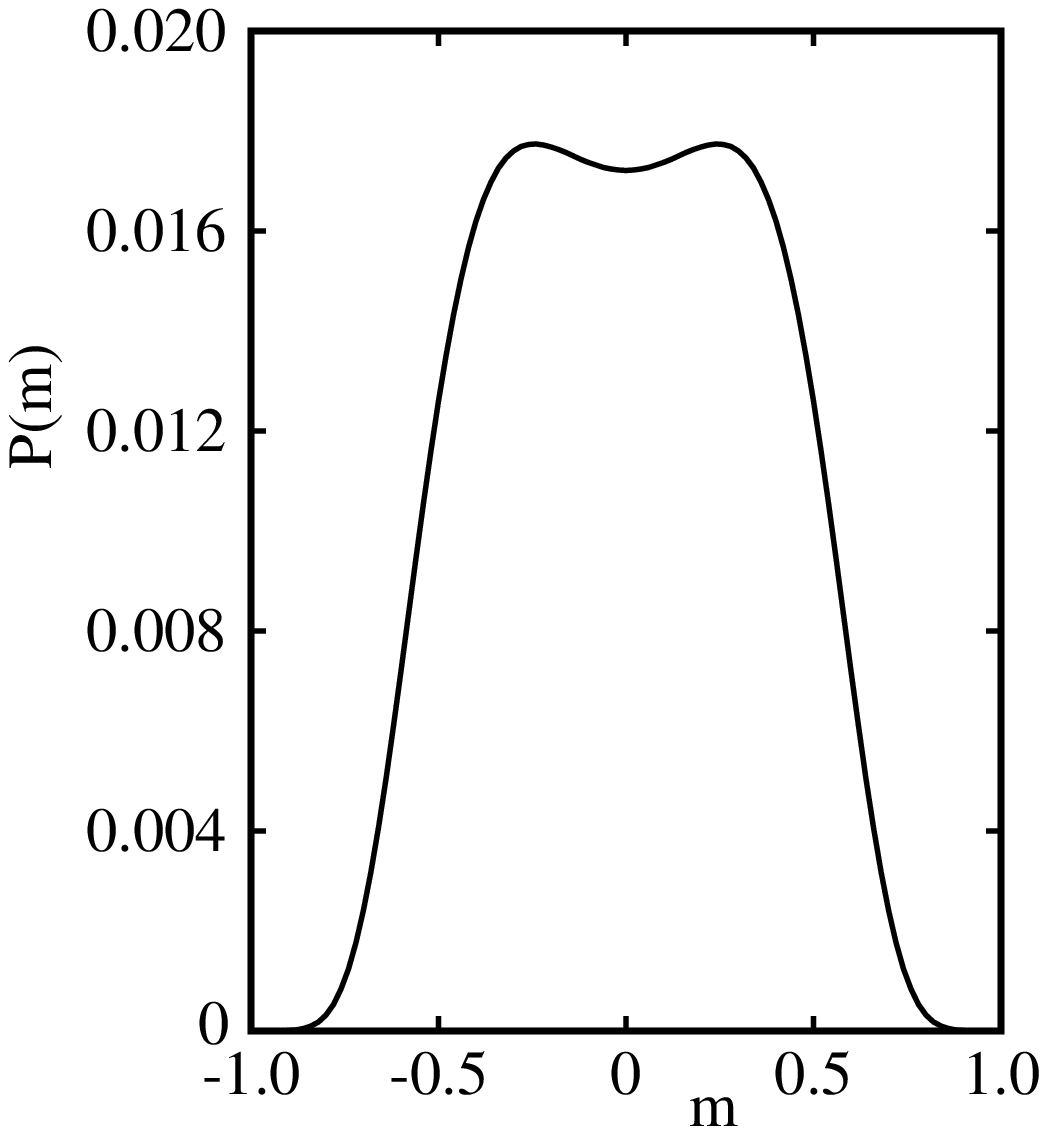}} \quad
\subfigure[]{\includegraphics[scale=0.6]{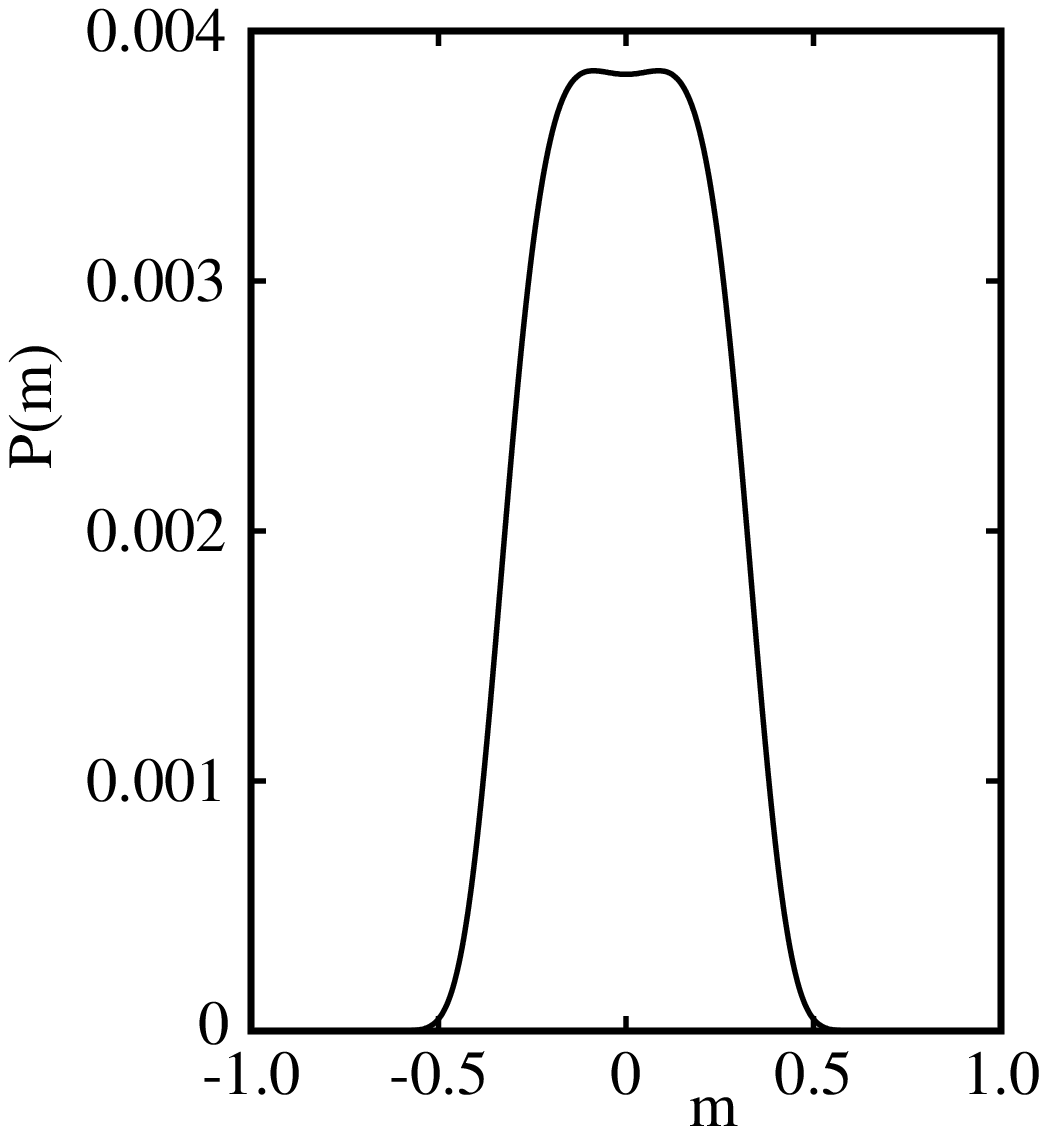}}
\vspace{-0.3cm}
\caption{\label{plotofP(M)}Plot of $P(m)$ versus $m$ as determined from Eqs.~\eqref{g(M)}--\eqref{E} at $T=\tci$ for (a) $N=100$ and (b) $N=800$. Note that $P(m)$ is symmetrical about $m=0$, and the maxima of $P(m)$ are at $|m|>0$. It is clear that $P(m)$ cannot be approximated by a Gaussian.}
\end{figure}

\begin{figure}[h!]
\centering
\includegraphics[scale=0.6]{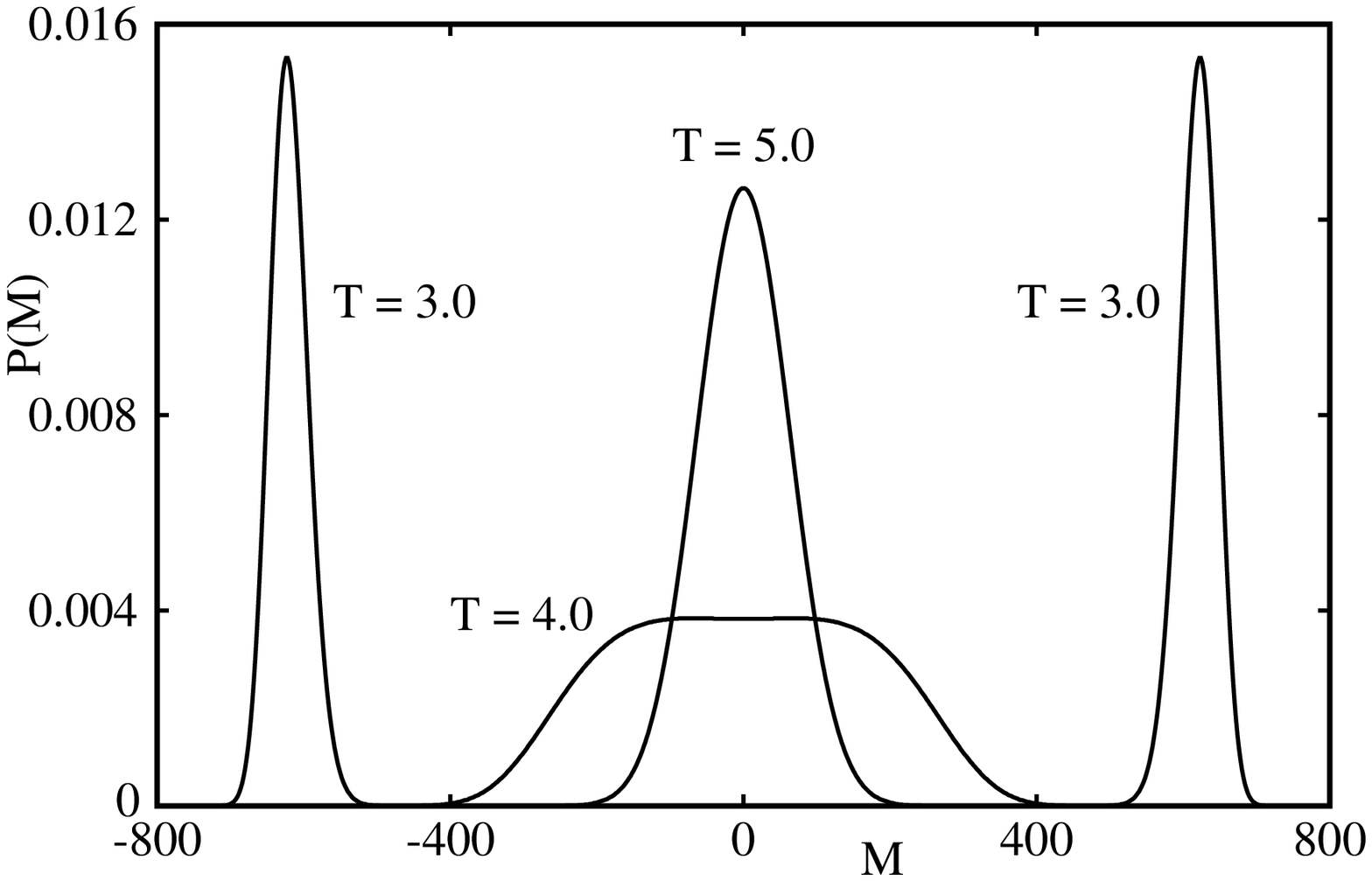}
\vspace{-0.2cm}
\caption{\label{proball}Plot of $P(M)$ at $T = 5$, $T=\tci=4$, and $T=3$ for $N=800$. As expected, $P(M)$ has a single maximum for $T>\tci$ and two maxima at $M \neq 0$ for $T < \tci$.}
\end{figure}

A plot of $P(m)$ for $N=100$ and $N=800$ as determined from Eqs.~\eqref{g(M)}--\eqref{E} at $T=\tci$ is given in Fig.~\ref{plotofP(M)}. Note that $P(m)$ is not a Gaussian and the maxima of $P(m)$ are at $|m| > 0$. To emphasize that the behavior of $P(m)$ at $T=\tci$ is qualitatively different than at other temperatures, we plot $P(M)$ for $T=3$, $T=\tci=4$, and $T=5$ in Fig.~\ref{proball}. We see that $P(M)$ has a single maximum for $T>\tci$ and has two maxima at $M \neq 0$ for $T < \tci$.

One way to characterize $P(m)$ is to compute the reduced fourth-order (Binder) cumulant, which is defined as~\cite{binder2}
\beq
U_4 = 1 - \frac{\overline{m^4}}{3 \overline{m^2}^2}. \label{cumulant}
\eeq
We use Eqs.~\eqref{g(M)}--\eqref{E} to compute $U_4$ and find that, as expected, $U_4 \approx 0$ for $T=5$, and hence $P(m)$ is well approximated by a single Gaussian for $T > \tci$ and large $N$. Similarly, for $T=3$ we find that $U_4 \approx 2/3$, which implies that $P(m)$ is well approximated as a sum of two Gaussians~\cite{landau}. At $T=\tci$ we find that $U_4 \approx 0.2706$ for $N=2 \times 10^7$, and hence at the critical temperature of the infinite system $P(m)$ is not well approximated by a Gaussian, even for large $N$. We also note that $U_4 \approx 0.4948$ at $T = \tcn$, the pseudocritical temperature at which the susceptibility for a given $N$ is a maximum.

\begin{figure}[htb]
\centering

\subfigure[\label{nn}]{\includegraphics[scale=0.7]{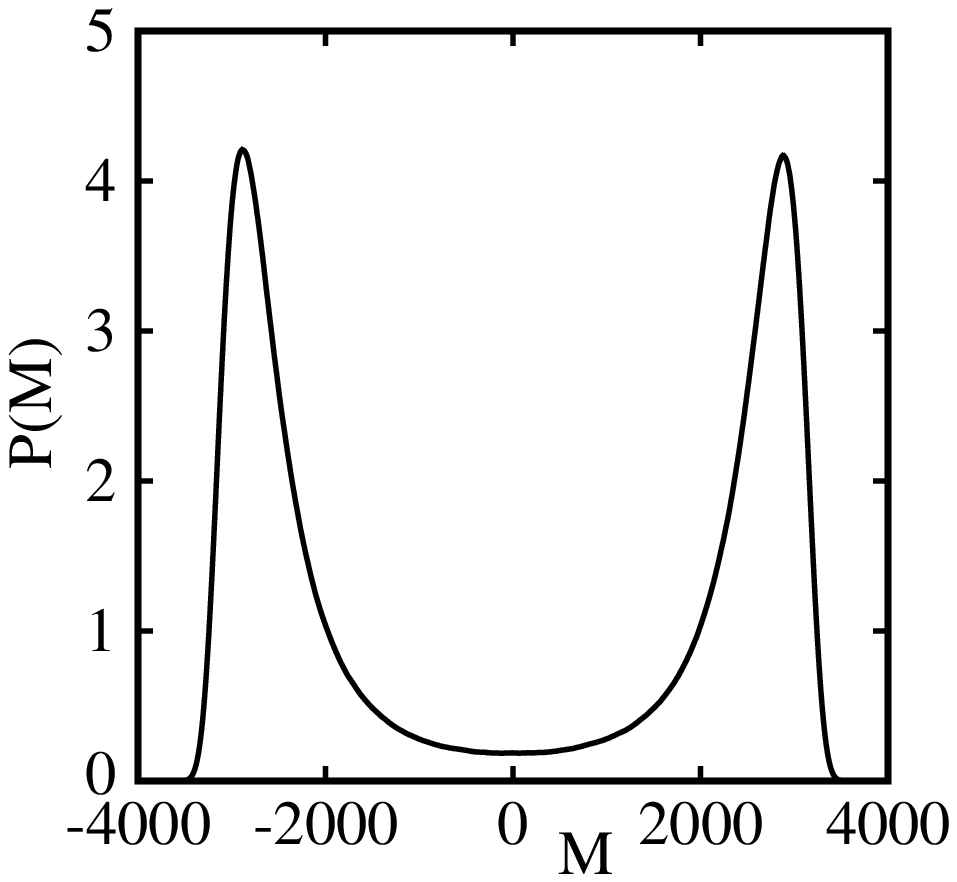}} \quad
\subfigure[\label{nnmmax}]{\includegraphics[scale=0.7]{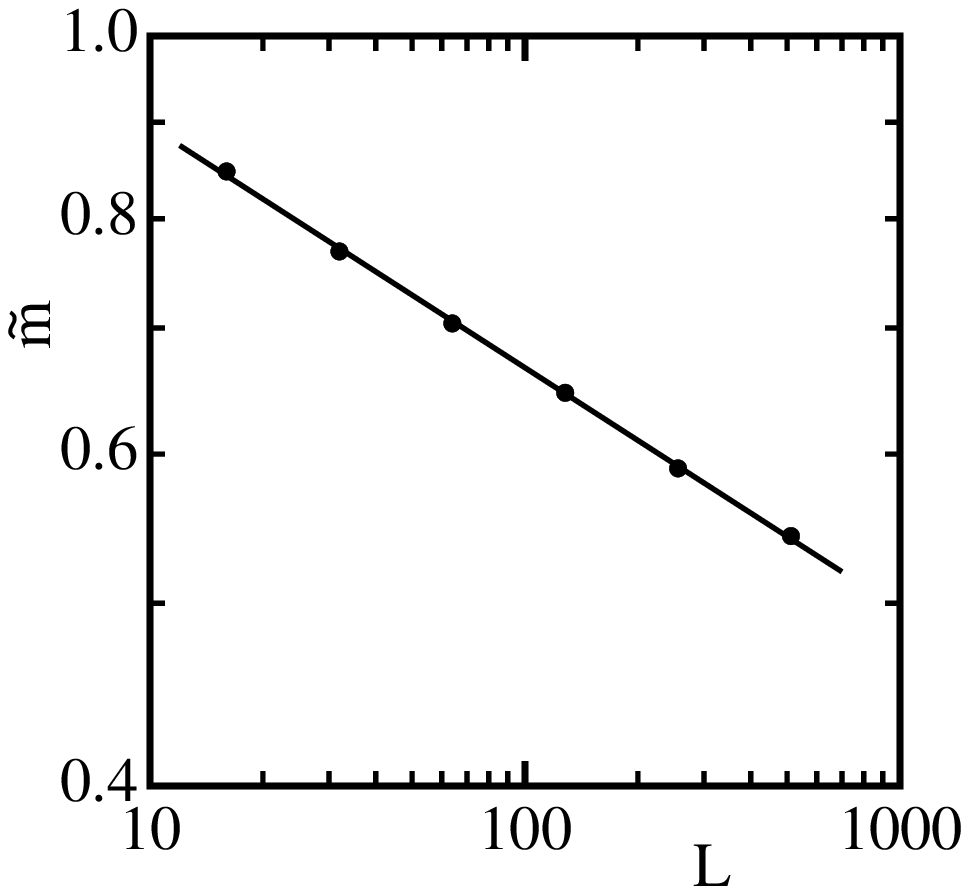}}
\caption{\mc\ results for the \nn\ \im\ at the critical temperature of the infinite system, $T_c=2/\ln (1 + \sqrt 2)$. (a) The probability $P(M)$ for linear dimension $L=64$ and $10^8$ \mc\ steps per spin. Note the existence of maxima at $M \approx \pm 2872$. (b) Log-log plot of the maxima of $P(m)$ for $m>0$ as a function of $L$. The slope is approximately $-0.128$, consistent with $\beta = 0.125$, the critical exponent for the mean magnetization.}
\end{figure}

It is interesting to compare the behavior of $\mt$ and $P(m)$ for the \fcim\ to their behavior in the \nn\ \im\ at the critical temperature of the infinite system. As shown in Fig.~\ref{nn}, the maxima of $P(m)$ for the \nn\ \im\ at $T = T_c=2/\ln (1 + \sqrt 2)$ as obtained from a \mc\ simulation are not at $m=0$~\cite{binder3}. However, the most probable and mean values of the magnetization both scale as $L^{-1/8}$ in two dimensions [see Fig.~\ref{nnmmax}], in contrast to their different scaling behavior in the \fcim. 

\section{\label{sec:restore}The Ginzburg parameter and the restoration of hyperscaling}

The different scaling behavior of the mean magnetization and the most probable magnetization in the \fcim\ implies that hyperscaling does not hold. As discussed in Ref.~\onlinecite{bigKlein}, hyperscaling is not satisfied by \mf\ theories, but hyperscaling is restored if the Ginzburg parameter, $G$, is held constant as the critical point is approached~\cite{bigKlein, tane}.

The definition of the Ginzburg parameter follows from the well known Ginzburg criterion for the applicability of \mft~\cite{criterion}. This criterion requires that the fluctuations of the order parameter be small compared to its mean value, that is, $G^{-1} = \xi^d \chi/\xi^{2d} \mbar^2 \ll 1$, where $\xi$ is the correlation length and $d$ is the spatial dimension. In the limit $G \to \infty$ the system is described exactly by \mf\ theory. The system is near-mean-field for $G \gg 1$ but finite.

To determine the dependence of $G$ on $N$ and $\e = |T-\tci|/\tci$, we use the \mf\ dependence of $\mbar$ and $\chi$ implied by the exponents in Eq.~\eqref{mfcpexp} and obtain $G = \xi^d \e^2$~\cite{bigKlein}. Because $N \sim \xi^d$, the Ginzburg parameter for the \fcim\ is given (up to a numerical constant) by
\beq
G = N \e^2. \label{G}
\eeq

We can show analytically that $\mt$ scales as $N^{-1/4}$ if $G$ is held constant. We substitute $T=\tci(1 +\e) $ in Eq.~\eqref{eq:approx}, assume that $\e = -(G/N)^{1/2}$ with $G$ a constant and $T < \tci$, and rewrite Eq.~\eqref{eq:approx} to leading order in $1/N$ as
\beq
-m - \frac{m^3}{3} + \frac{m}{1-(G/N)^{1/2}} = 0,
\label{eq:leadingorder}
\eeq
where $qJ/\tci=1$. If we let $[1 - (G/N)^{1/2}]^{-1} \approx 1 + (G/N)^{1/2}$, we obtain
\beq
\mt =3^{1/2} \Big(\frac{G}{N}\Big)^{\!1/4} \sim N^{-1/4}. \qquad \mbox{(constant Ginzburg parameter)}
\label{restore}
\eeq

From Eq.~\eqref{eq:approx} we see that the scaling of $\mt$ is determined by the way the coefficient of the linear term in $m$ vanishes. Instead of working at the critical temperature of the infinite system, we determine how $\mt$ scales with $N$ at the pseudocritical temperature $\tcn$. To that end we define
$\en = (\tci - \tcn)/\tci$,
and note that the coefficient of the linear term in Eq.~\eqref{eq:approx} can be written as
$\en + \tci/N\tcn$.

\begin{figure}[t]
\centering
\includegraphics[scale=0.5]{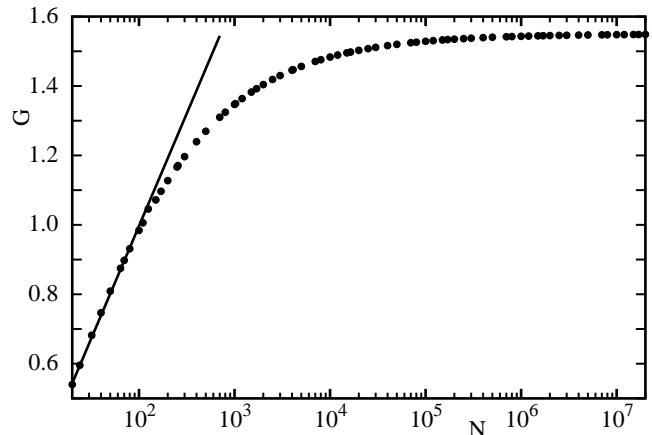}
\vspace{-0.4cm}
\caption{\label{fig:gn}Plot of the Ginzburg parameter $G = N \en^2$ with $\en = (\tci-\tcn)/\tci$ versus $\log N$. The pseudocritical temperature, $\tcn$, corresponds to the temperature at which $\chi$ is a maximum for a given $N$. The smallest value of $\en$ is $2.8 \times 10^{-4}$ for $N=2 \times 10^7$. A least squares fit of $G(N)$ for $N \leq 10^2$ yields $G(N) \approx -0.3000 + 0.6484 \log N$. The plot suggests that $G$ is increasing no faster than $\log N$ for larger values of $N$.}
\end{figure}

We can show that hyperscaling is apparently restored if \fss\ is done at $\tcn$~\cite{berche}. We compute $\tcn$ numerically using Eqs.~\eqref{g(M)}--\eqref{E}. The corresponding results for the Ginzburg parameter $G = N \en^2$ versus $\log N$ are shown in Fig.~\ref{fig:gn}. We see that $G$ is a slowly increasing function of $\log N$ for large $N$ and is increasing no faster than $\log N$ for large $N$. We were unable to fit the $N$-dependence of $G$ to a simple analytical form in the range $10^6 \leq N \leq 2 \times 10^7$ and were unable to distinguish between $G$ approaching a constant as $N \to \infty$ or $G$ increasing indefinitely, albeit less than logarithmically. This behavior is consistent with logarithmic corrections to the \mf\ behavior of quantities such as $\chi$ obtained by renormalization group calculations for the Ising model in four dimensions~\cite{berche, fisher}.

\begin{figure}[t]
\centering
\includegraphics[scale=0.6]{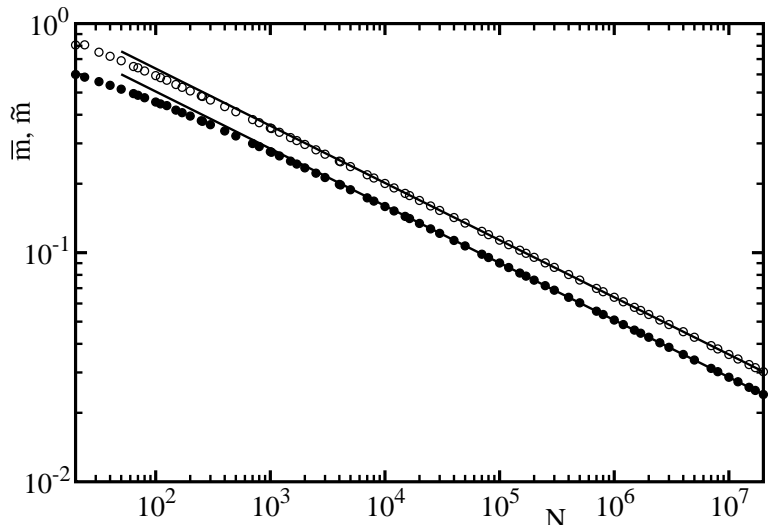}
\vspace{-0.4cm}
\caption{\label{fig:almostGconstant}Log-log plot of $\mt$ (open circles) and $\mbar$ (filled circles) versus $\log N$ for $N\leq 2 \times 10^7$ at the pseudocritical temperature corresponding to the values of $G$ shown in Fig.~\ref{fig:gn}. The slopes from a least squares fit in the range $10^5\leq N\leq2\times10^7$ of $\mt$ and $\mbar$ are $-0.2496$ and $-0.2494$, respectively, consistent with $\mt, \mbar \sim N^{-1/4}$, and the restoration of hyperscaling.}
\end{figure}

In Fig.~\ref{fig:almostGconstant} we show the $N$-dependence of $\mbar$ and $\mt$ computed at $T=\tcn$, the values of $T$ corresponding to the pseudocritical point. Least squares fits to $\mt$ and $\mbar$ yield slopes of $-0.2496$ and $-0.2494$, respectively, consistent with $\mt, \mbar \sim N^{-1/4}$, and the apparent restoration of hyperscaling if \fss\ is done at the pseudocritical temperature. We conclude that $G$ is increasing sufficiently slowly with $N$ for $N \leq 2 \times 10^7$ so that we cannot distinguish numerically between the results of constant $G$ or possible corrections.

We also investigated the $N$-dependence of the specific heat $C$ at both $\tcn$ and $\tci$. We find that $C$ is a slowly increasing function of $N$ at both temperatures, and we unable to fit the $N$-dependence of $C$ to a simple analytic function. Hence, we are unable to conclude if $C$ is approaching a constant as is predicted by \mft\ or if there are logarithmic corrections.

\section{\label{sec:anal}Analytical Calculation of the mean magnetization}

To calculate the scaling behavior of $\mbar$ at $T=\tci$, we expand $\ln P(m)$ in a Taylor series in $m-\mt$, where $\mt$ is the most probable value of $m$ as given by Eq.~\eqref{scaling of m2}. We have
\begin{eqnarray}
\ln P(m) &\approx& \ln P(\mt) + \half (m-\mt)^2 \frac{d^2 \ln P(m)}{dm^2}\bigg|_{m=\mt}\nonumber \\
&&{}+ \frac{1}{3!} (m-\mt)^3 \frac{d^3 \ln P(m)}{dm^3}\bigg|_{m=\mt} + \frac{1}{4!} (m-\mt)^4 \frac{d^4 \ln P(m)}{dm^4}\bigg|_{m=\mt}. \label{eq:4thorder}
\end{eqnarray}
In analogy to the form of the free energy in Landau-Ginzburg theory, we will need to keep terms only to fourth-order in $(m-\mt)^4$~\cite{machta}. We also expect that the second and third derivatives of $\ln P(m)$ to both approach zero as $N \to \infty$ and $(d^4 \ln P(m)/dm^4)_{m=\mt}$ to be independent of $N$.

We have to leading order in $1/N$ that
\begin{align}
\frac{d^2 \ln P(m)}{dm^2} & = -\frac{1}{1-m^2} + \frac{\beta q J}{1 - 1/N} + \frac{1}{N}\frac{1+m^2}{(1-m^2)^2} \label{d2lnP}, \\
\noalign{\noindent and hence}
\frac{d^2 \ln P}{dm^2}\Big|_{m=\mt} & \approx -1 - \mt^2 + 1 + \frac{1}{N} + \frac{1}{N} = -\frac{4}{N}. \label{d2lnPtc}
\end{align}
Note that $(d^2 \ln P/dm^2)_{m=\mt} < 0$, which is consistent with $\mt$ being the most probable value.

We also have to leading order that
\begin{equation}
\frac {d^3 \ln P}{dm^3} = -\frac{2m}{(1-m^2)^{2}} 
\quad \mbox{and} \quad
\frac {d^4 \ln P}{dm^4} = -\frac{2}{(1-m^2)^{2}}. 
\end{equation}
Hence to leading order in $1/N$ we have
\begin{equation}
\frac {d^3 \ln P}{dm^3} \bigg|_{m=\mt} \approx -2 \Big(\frac{6}{N}\Big)^{\!1/2} \quad \mbox{and} \quad 
\frac {d^4 \ln P}{dm^4}\bigg|_{m=\mt} \approx -2. \label{deriv4}
\end{equation}

We can interpret $\ln P(m)$ as the free energy per spin. Because $(d^2 \ln P(m)/dm^2)_{m=\mt}$ and $(d^3 \ln P(m)/dm^3)_{m=\mt}$ both go to zero as $N \to \infty$, we have from Eqs.~\eqref{eq:4thorder} and \eqref{deriv4} that~\cite{botet2}
\beq
\mbar = \frac{\int\limits_0^1 m \, e^{-N (m-\mt)^4/12}\,dm}{\int\limits_0^1 e^{-N(m-\mt)^4/12}\,dm} \qquad (N \gg 1).
\eeq
We change variables to $x=(m-\mt)(N/12)^{1/4}$ and keep only the leading order term in $N$. The upper limit of integration, $x_{\max}=(1 - \mt) (N/12)^{1/4} \sim N^{1/4} \to \infty$ as $N \to \infty$. Similarly, the lower limit of integration $x_{\min}=- \mt (N/12)^{1/4} \sim N^{-1/4} \to 0$ as $N \to \infty$. Hence, for large $N$ we obtain
\beq
\mbar = \Big(\frac{12}{N} \Big)^{\!1/4} \dfrac{\int\limits_{0}^\infty \! x \, e^{-x^4}\,dx}{\int\limits_{0}^\infty \! e^{-x^4}\,dx} \approx 0.91 N^{-1/4} . \label{eq:mbar}
\eeq
The leading correction to $\mbar$ in Eq.~\eqref{eq:mbar} is proportional to $N^{-1/2}$. Similar considerations yield the scaling behavior of $\chi$ given in Eq.~\eqref{chi}.

It is easy to check that $(d^n \ln P(m)/dm^n)_{m=\mt}$ for $n > 4$ is either independent of $N$ ($n$ even) or proportional to $N^{-1/2}$ ($n$ odd), thus justifying the assumption in Eq.~\eqref{eq:4thorder} that higher-order terms in the expansion of $\ln P(m)$ can be neglected.

The form of $\ln P(m)$ in Eq.~\eqref{eq:4thorder} can be used to compute the cumulant defined in Eq.~\eqref{cumulant}. The result is $U_4 \approx 0.271$ at $T=\tci$, which is consistent with the computed value of $U_4=0.2706$ using the exact density of states for $N=2 \times 10^7$.

\section{Scaling at the Spinodal}

\subsection{Simple scaling argument}
Because the spinodal is a line of critical points, we expect that finite size scaling at the Ising spinodal proceeds similarly to our analysis at the Ising \mf\ critical point. We assume that $T <\tci$ and vary the field $h$ near the spinodal field $h_s$. In terms of $\dh = (h-h_s)/h_s$ the usual scaling relations are~\cite{bigKlein}
\begin{align}
\psibar & \sim \dh^{1/2} \label{eq:psibar}\\
\chi & \sim \dh^{-1/2} \label{eq:chis} \\
\xi & \sim \dh^{-1/4}, \label{eq:xis}
\end{align}
where the order parameter $\psibar = \mbar - \ms$ is related to the mean magnetization per spin near the spinodal, and $\ms$ is the value of the magnetization at the spinodal. We use Eq.~\eqref{eq:xis} to obtain $\psibar \sim \xi^{-2}$ and $\chi \sim \xi^2$. If we assume the upper critical dimension to be six at the spinodal~\cite{unger}, we have $N \sim \xi^6$, and hence
\begin{align}
\psibar & \sim N^{-1/3} \label{eq:mnearspinodal} \\
\chi & \sim N^{1/3}. \label{eq:chisn}
\end{align}
We will derive these results in the following without assuming that
$N \sim \xi^6$ at the spinodal.

\subsection{Numerical results}
The numerical evaluation of the $N$-dependence of various quantities such as $\psibar$ and $\chi$ as a function of $N$ at $h=h_s$ using the exact density of states in Eq.~\eqref{g(M)} is more subtle than at the critical temperature because we must include only values of $M$ corresponding to the metastable state. To understand this restriction, imagine a \mc\ simulation of the \fcim\ at $T<\tci$ and magnetic field $h =h_0>0$. Because $h_0>0$, the values of $M$ are positive. After equilibrium has been reached, we let $h \to -h_0$. If $h_0$ is not too large, the system will remain in a metastable state for a reasonable number of \mc\ steps per spin. To compute $\chi$ associated with the pseudospinodal (the spinodal is defined only in the limit $N \to \infty$ for the \fcim), we must include only those values of $M$ that are representative of the metastable state. As discussed in Ref.~\onlinecite{zeros}, the values of $M$ that may be included in thermal averages of the metastable state must satisfy the condition that $M \gtrsim M_{\rm ip}$, where $M_{\rm ip}$ is the value of $M$ at the inflection point of $P(M)$. We set $d^2\ln P(M)/dM^2 = 0$ and use Eq.~\eqref{d2lnP} to find that~\cite{zeros}
\beq
M_{\rm ip} = \sqrt{N^2 \Big (1-\frac{1}{\beta q J}\Big) + \frac{N}{\beta q J}} \label{Mmin}.
\eeq

We follow Ref.~\cite{stauffer} and choose $T=4\tci/9 = 16/9$. Hence $\f = \beta q J = 9/4$ in Eq.~\eqref{Mmin}. For this value of $z$ we obtain $\hs \approx 1.2704$~\cite{hs}.

\begin{figure}[t]
\centering
\includegraphics[scale=0.6]{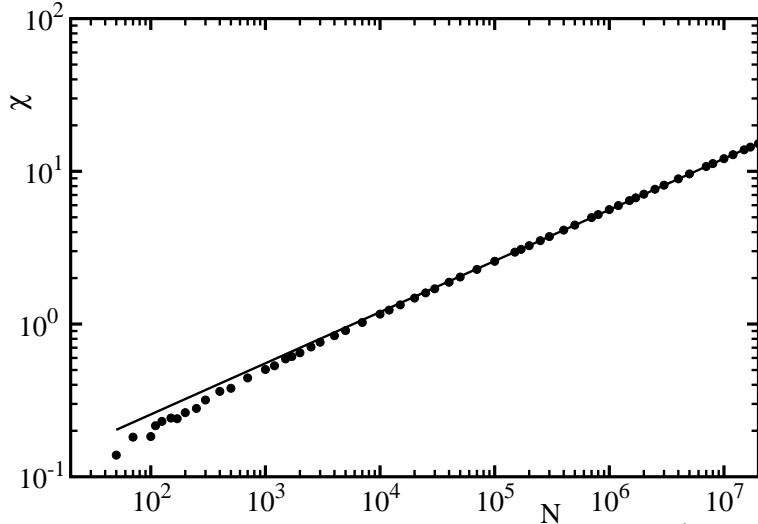}
\vspace{-0.5cm}
\caption{\label{chis}Log-log plot of $\chi$, the susceptibility per spin, versus $N$ at $h=h_s$ and $T=4 \tci/9$ using the exact density of states in Eq.~\eqref{g(M)} and the requirement that $M \geq M_{\rm ip}$. A least square fit for $10^5\leq N\leq 2\times10^7$ yields a slope of 0.335, consistent with the exponent 1/3 in Eq.~\eqref{eq:chisn}.}
\end{figure}

\begin{figure}[htb]
\centering
\subfigure[\label{mbarspinodal}]{\includegraphics[scale=0.45]{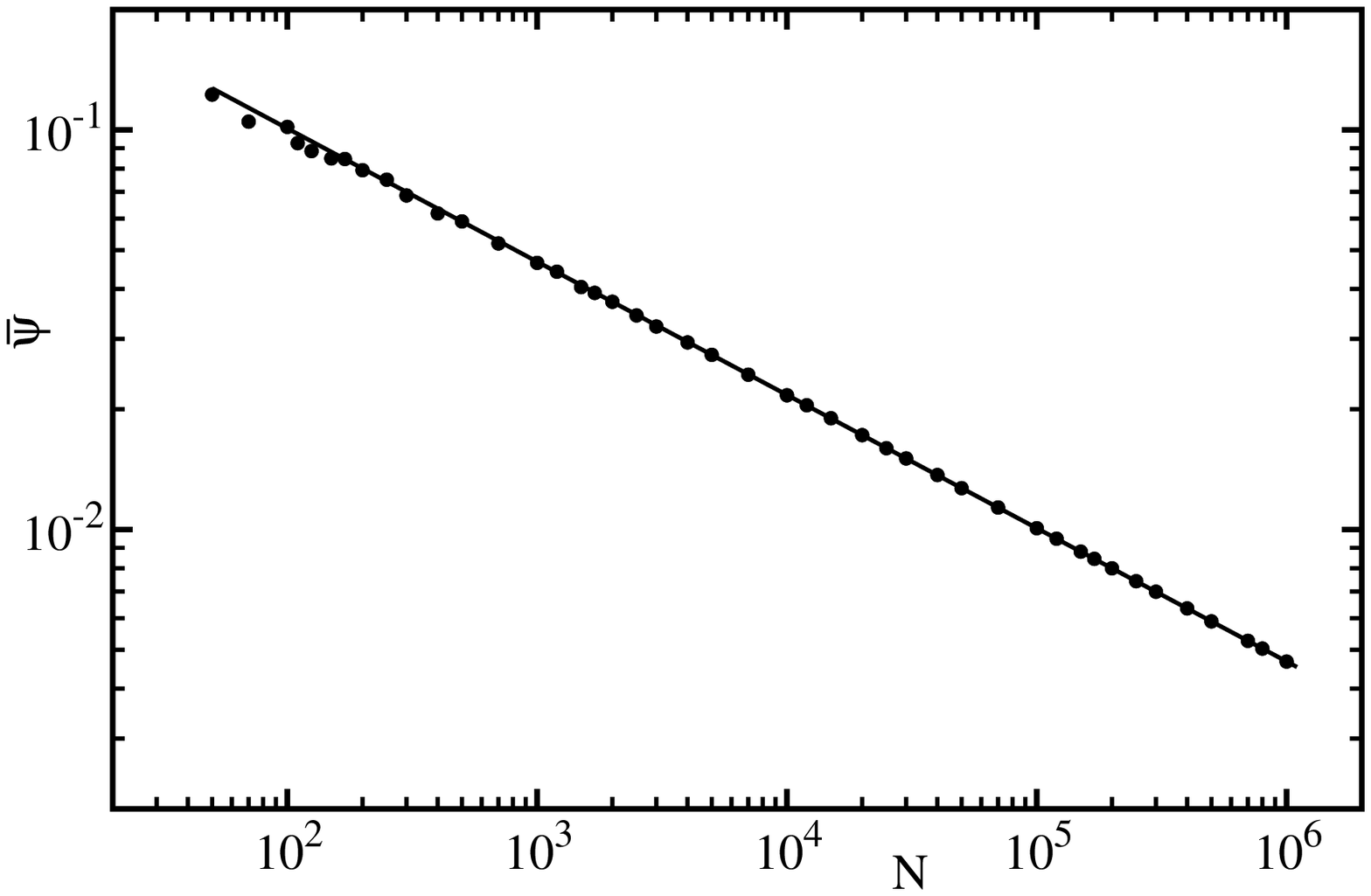}}
\subfigure[\label{mtspinodal}]{\includegraphics[scale=0.45]{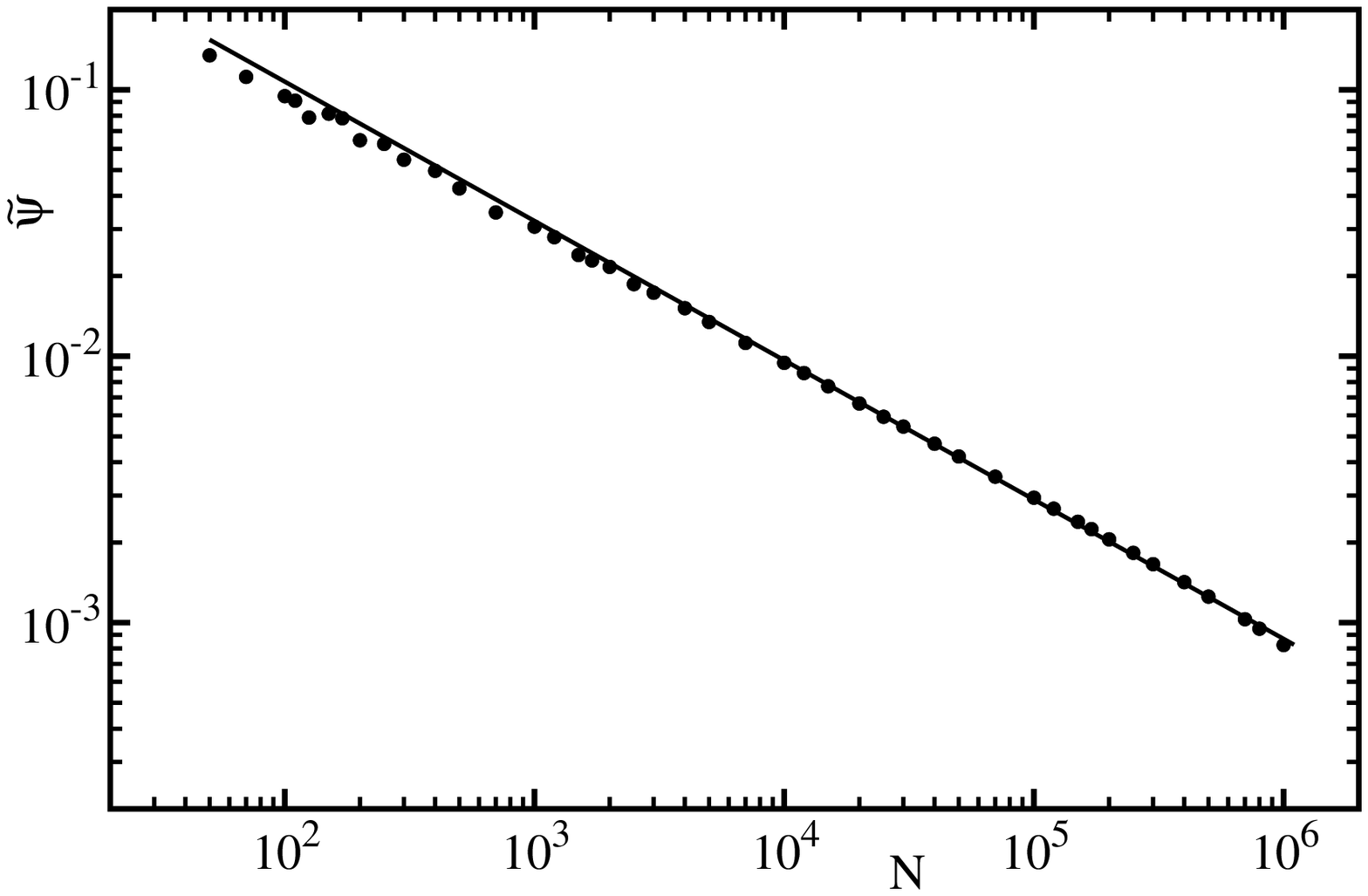}}
\caption{\label{mbars}(a) Log-log plot of the mean value of the order parameter, $\psibar$, versus $N$ at $h=h_s$ and $T=4\tci/9$. The slope is $\approx -0.334$, which is consistent with Eq.~\eqref{eq:mnearspinodal}. (b) Log-log plot of the most probable value of the order parameter, $\psit$, at $h=h_s$ and $T=4\tci/9$. The slope is $\approx - 0.523$, consistent with the exponent in Eq.~\eqref{msmost}. We see that the $N$-dependences of $\psibar$ and $\psit$ at the spinodal differ.}
\end{figure}

Our numerical results for $\chi$ at $h=h_s$ for increasing values of $N$ are shown in Fig.~\ref{chis} using the exact density of states in Eq.~\eqref{g(M)} and values of $M > M_{\rm ip}$. The slope of 0.335 is consistent with Eq.~\eqref{eq:chisn}. Similarly, we find that
a log-log plot of $\psibar$ versus $N$ yields a slope of $-0.334$ [see Fig.~\ref{mbarspinodal}] in agreement with Eq.~\eqref{eq:mnearspinodal}. A log-log plot of the most probable value of $m$ near the spinodal yields the scaling behavior [see Fig.~\ref{mtspinodal}]
\beq
\psit \sim N^{-1/2}. \label{msmost}
\eeq

\subsection{Analytical derivation}

The analytical calculation of the $N$-dependence of $\psibar$, $\psit$, and $\chi$ at the spinodal proceeds similarly to the derivation at the critical temperature. We can use Eq.~\eqref{first} with $N \to \infty$ to show that the value of $m$ at the spinodal is given by
$(1-\ms^2)^{-1} - q \beta J = 0$,
or
\beq
\ms = \sqrt{\frac{\beta q J - 1}{\beta q J}} = \sqrt{\frac{\f-1}{\f}}. \label{ms}
\eeq
The corresponding value of $h_s$ can be obtained by substituting $m=\ms$ into Eq.~\eqref{first} in the limit $N \to \infty$.

To find the leading correction to the most probable value of $m$ near the spinodal, we substitute $m = \ms + \psi$ in Eq.~\eqref{first} and assume that $\psi \ll \ms$ for $N \gg 1$. The result is
\begin{align}
\frac{d \ln P}{dm} & \approx \half \ln \frac{1-\ms}{1+\ms} + z \ms + \beta h_s - \frac{1}{1 - \ms^2} \psi + z \psi - \frac{\ms}{(1 -\ms^2)^2} \psi^2 \notag \\
& \quad + \frac1N \frac{\ms}{1 - \ms^2} + \frac1N \frac{1 + \ms^2}{(1 - \ms^2)^2} \psi + \frac{z\ms}{N} + \frac{z \psi}{N} = 0. \label{dP/dm}
\end{align}
The sum of the first three terms on the right-hand side is zero. We will assume that $\psi \sim N^{-1/2}$ and determine if this assumption is consistent with the solution to Eq.~\eqref{dP/dm}.

The terms proportional to $N^{-1/2}$ are
\beq
\Big [-\frac{1}{1-\ms^2} + \f \Big ]\psi,
\eeq
which sum to zero using Eq.~\eqref{ms}. The terms proportional to $N^{-1}$ include
\beq
-\frac{\ms}{(1 - \ms^2)^2} \psi^2 + \frac1N \frac{\ms}{1 - \ms^2} + \frac{z \ms}{N},
\eeq
which also must sum to zero. The result for $\psi^2$ to order $1/N$ is
\beq
\psi^2 = \frac{(1 - \ms^2)^2}{N}\Big[\frac{1}{1-\ms^2} + z \Big]= \frac{2}{Nz}. \label{psi2}
\eeq
The quantity $\psi$ in Eq.~\eqref{psi2} represents the most probable value, which we write in the following as $\psit$. Hence, we conclude that $\psit \sim N^{-1/2}$, in agreement with the numerical result in Eq.~\eqref{msmost} and the slope in Fig.~\ref{mtspinodal}.

Near the spinodal the Ginzburg parameter $G_s$ is given by $G_s=\xi^d \psibar^2/\chi \sim N \dh^{3/2}$~\cite{bigKlein}, where we have used Eqs.~\eqref{eq:xis} and \eqref{eq:chisn}. In analogy to our discussion in Sec.~\ref{sec:restore}, we can show that $\psit \sim N^{-1/3}$ if $G_s$ is held fixed as $\dh$ is varied at constant temperature.

Similarly, we find for large $N$ at $T=4\tci/9$ and $h=\hs$ that
\begin{align}
\frac{d^2 \ln P}{dm^2} & = -\frac{2\ms}{(1 - \ms^2)^2}\psi \sim N^{-1/2}, \\
\noalign{\noindent and}
\frac {d^3 \ln P}{dm^3} & = -\frac{2\ms}{(1-\ms^2)^3} \sim N^0.
\end{align}
We see that $d^2 \ln P/dm^2 \sim N^{-1/2}$ and $d^3 \ln P/dm^3$ is independent of $N$ in the limit $N \to \infty$. Hence, we can show that $\psibar \sim N^{-1/3}$ and $\chi\sim N^{1/3}$ at the spinodal in agreement with Eqs.~\eqref{eq:mnearspinodal} and \eqref{eq:chisn}.

\section{Discussion}

We have shown that \fss\ done at $\tci$, the critical temperature of the \fcim\ in the limit $N \to \infty$, gives results that differ from our usual understanding of \mf\ systems. In addition, we found that \fss\ yields different results depending on how the \mf\ limit is approached.

In particular, the Gaussian approximation often associated with \mft\ does not hold at $T=\tci$, and the probability distribution of the magnetization is not a Gaussian, even in the limit $N \to \infty$. Also our results are inconsistent with the assumption that the scaling properties of the \fcim\ at the critical temperature of the infinite system are the same as the scaling properties of the \nn\ Ising model (when $N$ is used as the scaling parameter) at the upper critical dimension, where hyperscaling is satisfied and the Ginzburg parameter is independent of the distance from the critical point and the spinodal.

The reason that the most probable value of the magnetization, $\mt$, and the mean value, $\mbar$, scale differently with $N$ at $T=\tci$ is that hyperscaling is not satisfied. However, the breakdown of hyperscaling does not affect the values of thermodynamic exponents such as $\beta$ and $\gamma$~\cite{note}.
In contrast, the most probable value of the magnetization is not a thermodynamic quantity and is affected. The breakdown of hyperscaling is consistent with results above the upper critical dimension where hyperscaling also does not hold if \fss\ is done at the critical point of the infinite system~\cite{berche}. To do \fss\ so that hyperscaling is restored, it is necessary to keep the Ginzburg parameter constant as $N$ is increased. It is also possible to do \fss\ at the pseudocritical temperature where the susceptibility is a maximum. In this case the Ginzburg parameter is either a constant or diverges more slowly than $\ln N$ for large $N$. Whether the latter dependence maintains hyperscaling or leads to logarithmic corrections cannot be determined from our numerical results.

To understand the different scaling behavior at $T=\tci$ and $T=\tcn$, we return to Eq.~\eqref{d2lnPtc} and interpret $\e$ as the coefficient of the quadratic term in the free energy. Hence at $T=\tci$ we have
\beq
\ei \sim -\frac{4}{N},\label{effeep}
\eeq
and the Ginzburg parameter $G = N \ei^2 \sim 1/N$, leading to $\mt \sim N^{-1/2}$. In contrast, if $G$ is held constant, $\ei \sim N^{-1/2}$, leading to $\mt \sim N^{-1/4}$. As shown in Fig.~\ref{fig:gn}, $G$ appears to increase more slowly than $\ln N$ for large $N$ if \fss\ is done at the pseudocritical temperature $T=\tcn$, the temperature corresponding to the maximum of the susceptibility. Because logarithmic corrections do not change the scaling laws~\cite{stanley}, we expect that corrections that are weaker than logarithmic will not affect the scaling of the most probable value of the magnetization. Hence, we conclude that $\mt \sim N^{-1/4}$ if \fss\ is done at $T=\tcn$.

It is remarkable that the \fcim, which is discussed in some undergraduate textbooks because of its simplicity~\cite{fcim1}, still yields surprises. In particular, the behavior of the \fcim\ at the critical point differs from that of the long-range \im\ with the Kac form of the interaction. This conclusion is not surprising because the interaction between spins in the \fcim\ does not have the Kac form for which \mf\ theory has been shown to be exact if the thermodynamic limit is taken \textit{before} the range of the interaction is taken to infinity.

Our results are a reminder that the applicability of \mf\ theories is subtle. A recent example is found in Ref.~\cite{defects}, where it was shown that the divergence of the specific heat of the long-range \im\ in one and two dimensions is neither \mf\ nor has the exponents associated with the \nn\ Ising model. We also note that experiments in systems that are well approximated by \mft\ are not usually done at fixed Ginzburg parameter. Hence, the interpretation of experimental results for such systems should be done with caution. 

\begin{acknowledgments}
We thank Jon Machta and Jan Tobochnik for useful discussions and Ammar Tareen for preliminary work on the \fcim. We also thank Julien Vidal and Lo\"ic Turban for bringing Refs.~\onlinecite{botet, botet2, shore} to our attention and the anonymous referees for useful comments. 
W.K.\ was supported by the DOE BES under grant number DE-FG02-95ER14498.

\end{acknowledgments}


\begin{thebibliography}{33}
\bibitem{botetnote} References~\onlinecite{botet} and \onlinecite{botet2} show that if a thermodynamic quantity of a \fc\ system scales as $(T-\tci)^{a_{\rm MF}}$, then it scales with $N$ at $T=\tci$ as $N^{-a_{\rm MF}/\nu_{\rm MF} d_c}$, where $\nu_{\rm MF}$ is the \mf\ critical exponent associated with the correlation length and $d_c$ is the upper critical dimension of the corresponding short-range system. They apply this prediction to the \fcim\ for which $\nu_{\rm MF} = 1/2$ and obtain the scaling results in Eqs.~\eqref{m} and \eqref{chi}. They also apply their argument to the \fc\ spin 1/2 Heisenberg model and the \fc\ 
quantum XY model in a transverse field.
\end{thebibliography}
\end{document}